\documentclass[aps,prb,twocolumn,superscriptaddress,usenames,dvipsnames,floatfix,10pt]{revtex4-2}
\usepackage{dcolumn}
\usepackage{bm}
\usepackage[T1]{fontenc}
\usepackage{float}
\usepackage{slashed}
\usepackage{amsmath}
\usepackage[pdftex]{graphicx}
\usepackage{epstopdf}
\usepackage{footmisc}
\usepackage{hyperref}

\usepackage[capitalize]{cleveref}

\hypersetup{colorlinks=true, linkcolor=blue, citecolor=red, urlcolor=magenta, pdftitle={amisi}, pdfauthor={S. Amisi}}

\begin{document}

\title{The hidden ferroelectric chiral ground state of silver niobate}

\author{Safari Amisi}
\thanks{These authors contributed equally to this work.}
\affiliation{Laboratoire de Physique des Solides et des Interfaces, Institut Sup\'e{}rieur P\'e{}dagogique de Bukavu, Democratic Republic of the Congo}
\affiliation{Theoretical Materials Physics, Q-MAT, University of Li\`e{}ge, B-4000 Sart Tilman, Belgium}
\author{Fernando G\'{o}mez-Ortiz}
\thanks{These authors contributed equally to this work.}
\affiliation{Theoretical Materials Physics, Q-MAT, University of Li\`e{}ge, B-4000 Sart Tilman, Belgium}

\author{Eric Bousquet}
\affiliation{Theoretical Materials Physics, Q-MAT,  University of Li\`e{}ge, B-4000 Sart Tilman, Belgium}

\author{Philippe Ghosez}
\affiliation{Theoretical Materials Physics, Q-MAT, University of Li\`e{}ge, B-4000 Sart Tilman, Belgium}

\begin{abstract}
Silver niobate is a conventional perovskite oxide compound, known to exhibit a rich polymorphism. Although often classified as antiferroelectric, its low-temperature structure remains unclear. Here, first-principles calculations reveal a previously overlooked and unusual rhombohedral ferroelectric phase with $R3$ symmetry that emerges as the thermodynamic ground state despite its almost degenerate energy and close energetic competition with previously proposed structures.
Remarkably, this phase is structurally chiral, with chirality emerging improperly from the coupling between polarization and in-phase rotations of the oxygen octahedra along [111], producing a ferri-chiral state with incomplete cancellation of local chiral motifs.
As a consequence, the phase exhibits significant natural optical activity comparable to that of quartz. Although energetically favored, its experimental observation may be hindered by kinetic limitations, potentially contributing to the ongoing controversy surrounding the low-temperature structure of silver niobate.
\end{abstract}

\maketitle 

\section{Introduction}

Antiferroelectric materials constitute a class of functional materials that exhibit a non-polar phase, but that can be turned ferroelectric under the application of an experimentally achievable external electric field. As such, the hallmark of antiferroelectrics is a flat energy landscape showing competing polar and non-polar minima separated by low energy barriers~\cite{{Amisi1},{rabeafe},{Aramberri2021}}.

Silver niobate, AgNbO$_3$, is often classified as one of the few known antiferroelectric perovskites \cite{Catalan2026}. Consistently, it exhibits a rich polymorphism similar to that of sodium niobate, NaNbO$_3$~\cite{safari}. 
At high temperature, it adopts the reference cubic perovskite structure ({\it Pm$\bar{3}$m}). Then, upon cooling, it shows a complex sequence of consecutive structural phase transitions to an antiferroelectric $Pbcm$ phase, as reported by several experimental~\cite{{Francombe},{kania98},{ratukan},{lukaszewski},{sciau},{zhangzhi},{niransir}} and theoretical studies~\cite{{ratukan},{niransir},{shigwada},{Moriwake12},{niranast},{CabSim}}: $ Pm\bar{3}m\xrightarrow{852K} P4/mbm \xrightarrow{660K} Cmcm  \xrightarrow{634K} Pmmn \xrightarrow{626K} Pbcm$. 
However, the low-temperature part of its phase diagram remains a subject of debate~\cite{{lukaszewski},{sciau},{fabrpet}}. Combining neutron and synchrotron powder diffraction, it was proposed that a ferroelectric {\it Pmc$2_1$} phase could coexist with the antiferroelectric $Pbnm$ phase (340$-$240 K)~\cite{{Moriwake},{tzhang},{yashima},{changszyf}}. This is made plausible by the fact that the {\it Pbcm} and {\it Pmc$2_1$} phases are structurally and energetically close and their coexistence could explain the weak ferroelectricity of the $Pbcm$ phase~\cite{jingSJ}. 
Independently, Zhang {\it et al.}~\cite{tzhang}, by examining the dielectric, pyroelectric, and ferroelectric data, proposed another ferroelectric phase below 250 K, which they identified as {\it R$3$c}, in line with an earlier proposal by Shigemi {\it et al.}~\cite{shigwada}.

Motivated by this ongoing controversy, we systematically re-explore the energy landscape of AgNbO$_3$ from first-principles. Our calculations reveal a so far hidden rhombohedral ferroelectric ground state with {\it R$3$} symmetry that is energetically favored over all previously proposed structures. Notably, this phase is not only ferroelectric but also chiral. Its chirality naturally arises from the combination of a polarization along the $[111]$ direction with an unusual $a^+a^+a^+$ pattern of rotation of the oxygen octahedra.

During the preparation of this manuscript, an independent first-principles study predicted the same rhombohedral ferroelectric chiral {\it R$3$} phase in AgNbO$_3$~\cite{song2026}.
While fully consistent with that independent study, our work further discuss the energetic hierarchy of competing distortions through a detailed symmetry-mode analysis and identify the {\it{R$3$}} structure as the potential thermodynamical ground state of AgNbO$_3$. Moreover, it elucidates the improper origin of its structural chirality through the coupling with non-chiral lattice modes and quantify the resulting natural optical activity, placing AgNbO$_3$ in the context of established optically active materials such as SiO$_2$.

\section{Computational details}\label{sec:technicaldeta}
All calculations were carried out using the {\sc abinit} density functional theory (DFT) package~\cite{abinit1,abinit2}, a planewave pseudopotential approach using the optimized norm-conserving pseudopotentials (ONCVPSP)~\cite{marques,hamann}, available on  {\sc pseudo dojo} server~\cite{pseudodojo}. All of the present ab initio calculations were performed using our optimized theoretical lattice constants. The Perdew-Burke-Ernzerhof  revised for solids functional (GGA$-$PBEsol)~\cite{perdew} was mainly employed, and, for comparison, some results have been checked at the local density approximation (LDA) level using Teter's extended norm conserving pseudopotentials~\cite{perdew2}.  The valence states for the computations are 2$s^2$2$p^6$3$s^1$  for Ag, 4$s^2$4$p^6$4$d^4$5$s^1$ for Nb and 2$s^2$2$p^4$ for O. Convergence was achieved for an energy cutoff of 45 hartrees for the plane-wave expansion (for both types of pseudopotentials) and a 8$\times$8$\times$8 grid of {\it k}-points for the Brillouin zone sampling of the single perovskite 5-atom cell. When condensing the AFD instabilities, we considered either for {\it I$4$/mcm, P$4$/mbm, Imma, Pnma} and  {\it Pmc$2_1-20$} phases, a 20-atom supercell corresponding to $\sqrt{2}a_0$, $\sqrt{2}a_0$, 2$a_0$, with a sampling of 6$\times$6$\times$4 {\it k}-points or, for the {\it I$4$/mmm}, {\it Cmcm}, {\it Pmmn}, {\it R$\bar{3}$c}, {\it R$3$c}, {\it Im$\bar{3}$} and  {\it R$3$}  structures, a 40-atom supercell corresponding to 2$a_0$, 2$a_0$, 2$a_0$ with  a sampling of 4$\times$4$\times$4 {\it k}-points. The  AFE/AFD {\it Pbcm} and polar {\it Pmc$2_1$}-40  phases were relaxed in a 40-atom supercell corresponding to $\sqrt{2}a_0$, $\sqrt{2}a_0$, 4$a_0$ and a sampling of 6$\times$6$\times$2 {\it k}-points. 
 We explicitly checked that the relative energy of the different phases is well converged and independent of the choice of the supercell. Structural relaxations were performed until the forces and stresses were smaller than 10$^{-7}$ hartrees/bohr and  10$^{-7}$ hartrees/bohr$^{3}$ respectively. The phonon frequencies, Born effective charges, and electronic dielectric tensor were calculated according to density functional perturbation theory (DFPT)~\cite{gonze}. The phonon dispersion curves where interpolated from the dynamical matrices on a  $ 2 \times 2 \times 2$ BCC grid of $q$-points, while treating explicitly the dipole-dipole interactions. 
For the calculation of the optical activity we used the linear-response approach grounded in density-functional perturbation theory implemented on {\sc{abinit}}~\cite{Zabalo-23}. We focus on the low-frequency limit of $\bar{\rho}(\omega)=\rho(\omega)/(\hbar \omega)^2$,
which tends to a constant as $\omega\rightarrow 0$, where $\rho(\omega)$ is the rotatory power and $\hbar$ the reduced Planck constant. 
The space group symmetry were checked by {\sc findsym} program~\cite{findsym}. To analyse  the group theory and relative contributions of different phonon modes to the distortions, we employed {\sc isotropy} and {\sc amplimode}  codes of the Crystallographic Bilbao server~\cite{isotropycode,amplimodecode,isodistortcode}.  

\section{Reference structure}\label{sec:cubicstruct}
We begin by considering the high-temperature cubic perovskite structure of AgNbO$_3$, for which symmetry leaves the lattice parameter as the only structural degree of freedom. Its relaxed value  is $a_0 = 3.948$~\AA, which lies between the lattice parameters of NaNbO$_3$ (AFE) and KNbO$_3$ (FE), and is consistent with experiment ($a_0 = 3.958$~\AA \cite{changszyf}) and previous theoretical results~\cite{{prosandeev}}.

Having relaxed the reference cubic structure, we next examine its Born effective charges, which provide information on the underlying chemistry of the system: Nb and O atoms exhibit strongly anomalous values ($Z^*_B=9.79,~Z^*_{O_\|}=-7.64$) reflecting the strong sensitivity of the hybridization between Nb 5{\it d} and O 2{\it p} orbitals under atomic displacement and driving a dominantly B-site polarization, while the charge of A-site cations (Ag) remains close to its nominal value ($Z^*_A=1.58$).
Regarding its electronic structure, AgNbO$_3$ appears as an indirect bandgap semiconductor with a calculated gap of 1.56 eV, comparable to NaNbO$_3$ (1.65 eV) and KNbO$_3$ (1.53 eV)  and underestimating the experimental gap of 2.8 eV, as commonly observed in DFT~\cite{{sciau}}.

\begin{figure}[htbp!]
\centering
\begin{tabular}{cc}
{\small (a)} & \includegraphics[width=7.5cm,keepaspectratio=true]{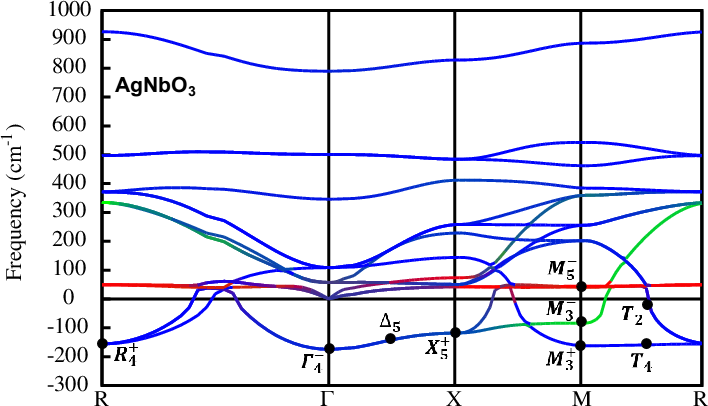} \\
{\small (b)} & \includegraphics[width=7.5cm,keepaspectratio=true]{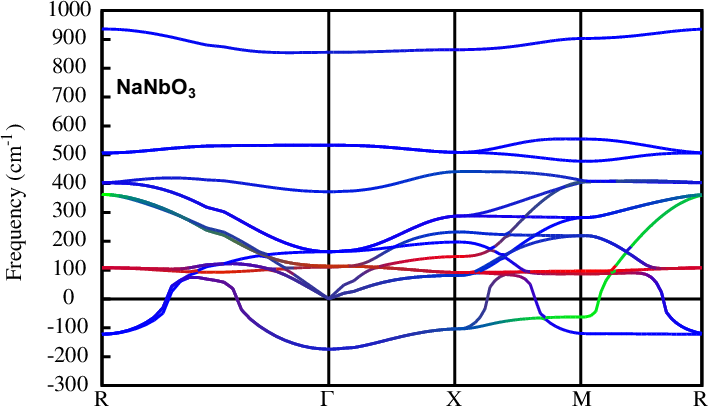} 
\end{tabular}
\caption{{\small (Color online) Calculated phonon dispersion curves of the cubic phase of (a)  AgNbO$_3$ and (b) NaNbO$_3$ (adapted from Ref.\cite{safari}), at the optimized volume. A color is assigned along  each line, according to the contribution of each kind of atom to the associated eigendsplacement vector (red for the Ag/Na atom, green for the Nb atom, and blue for O atoms). The labels of main soft modes are highlighted in panel (a).}}
\label{fig:figure1}
\end{figure}

To gain a global view of the lattice instabilities that may drive structural phase transitions, we next compute the phonon dispersion curves of the high-symmetry cubic phase of AgNbO$_3$ (Fig.~\ref{fig:figure1}a). These curves appear closely similar to those of NaNbO$_3$ (Fig.~\ref{fig:figure1}b)~\cite{safari}.
According to their Goldschmidt tolerance factor ($t \lesssim 1$), these compounds combine two types of unstable branches. 
On the one hand, they show an unstable branch associated to polar/antipolar motions, which is relatively flat in the $\Gamma-X-M$ plane. 
It is reminiscent of what happens in KNbO$_3$ and  indicative of a chain-like ferroelectric instability in real space, with weak inter-chain interactions~\cite{Yu-95, Ghosez-98}. On the other hand, and distinctly from KNbO$_3$, they show another unstable branch related to antiferrodistortive rotations of the oxygen octahedra with very little dispersion from $R$ to $M$.  

All these instabilities have very similar amplitudes and appear very competitive. Quantitatively, their magnitudes and relative strengths differ slightly from one compound to another, providing a microscopic explanation for their distinct low-temperature phases.
These quantitative differences are  summarized in Table~\ref{tab:table2}, in line with what is illustrated in Fig.~\ref{fig:figure1}. Our mode labels follow the convention of Miller and Love with the origin at the Nb site; the corresponding labels for the Ag-centered setting are provided in Appendix \ref{sec:phononlabels}. The rich set of instabilities highlighted here provides a roadmap for exploring how the various modes condense, which we examine in detail in the following section to understand the low-temperature phases of AgNbO$_3$.
 
\begin{table}[htbp!]
\renewcommand{\arraystretch}{1.35}
\centering
\caption{{\small Symmetry label, atoms involved, type and frequency $\omega$ (cm$^{-1}$) of the unstable phonon modes at selected q-points of the Brillouin zone in the high symmetry cubic phase of AgNbO$_3$ and NaNbO$_3$. FE, AP and AFD refer respectively to ferroelectric, anti-polar and antiferrodistortive character.}}
\begin{tabular}{lcllcc}
\hline                                             %
\hline
\multicolumn{2}{c}{{\it q}-points}              & Mode (atoms)   & Type    &\multicolumn{2}{c}{$\omega$ (cm$^{-1}$)} \\ 
         &                                      &                      &   &AgNbO$_3$    & NaNbO$_3$~\cite{safari}\\\hline
$\mathit{\Gamma}$ & (0, 0, 0)                   & $\mathit{\Gamma}_4^-$ (Ag, Nb, O) &FE     & 175{\it i}  & 174{\it i} \\
$\Delta$ &(0, $\frac{1}{4}$, 0)                 & $\Delta_5$ (Ag, Nb, O) &AP      & 134{\it i}   &129{\it i} \\
$X$      & ($\frac{1}{2}$, 0, 0)                & $X_5^+$ (Nb, O)       &AP & 118{\it i}   &103{\it i} \\
$M$      & ($\frac{1}{2}$, $\frac{1}{2}$, 0)    & $M_3^+$ (O)             &AFD    & 162{\it i}  & 120{\it i}\\
         &                                      & $M_3^-$ (Nb, O)              &AP    & 84{\it i}    & 62{\it i}\\
$T$      &($\frac{1}{2}$, $\frac{1}{2}$, $\frac{1}{4}$)& $T_4$ (Ag, Nb, O)        &AFD/AP & 159{\it i}  & 121{\it i}\\
$R$      &($\frac{1}{2}$, $\frac{1}{2}$, $\frac{1}{2}$)& $R_4^+$  (O)   &AFD & 156{\it i} & 122{\it i} \\
\hline
\hline
\end{tabular}
\label{tab:table2}
\end{table}
\begin{figure*}[htbp!]
\centering
\begin{tabular}{cc}
 \includegraphics[width=6.5cm,keepaspectratio=true]{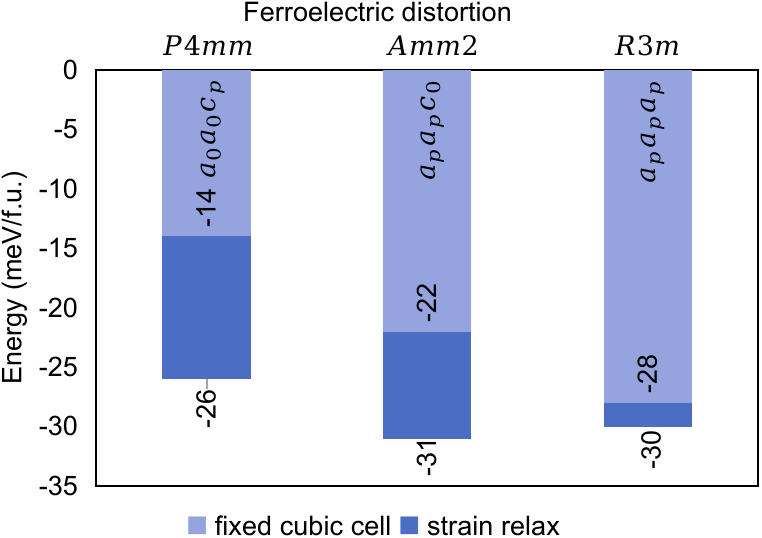} & \includegraphics[width=9cm,keepaspectratio=true]{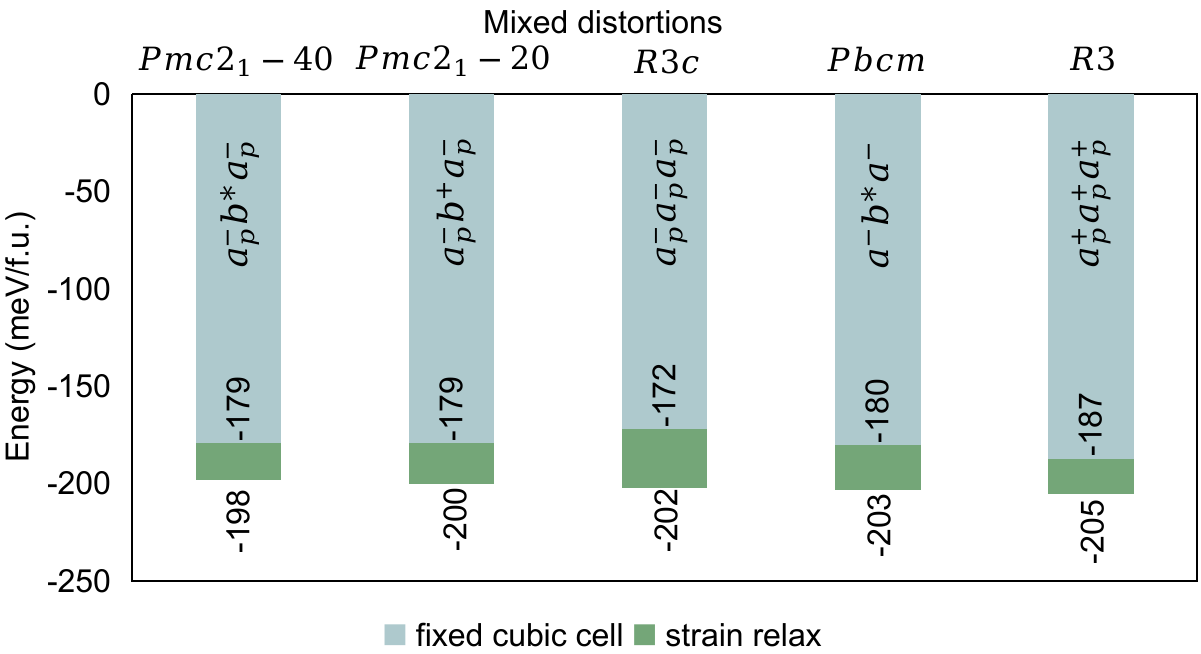} \\
 {\small (a)} & {\small (c)}\\
\multicolumn{2}{c}{\includegraphics[width=13.2cm,keepaspectratio=true]{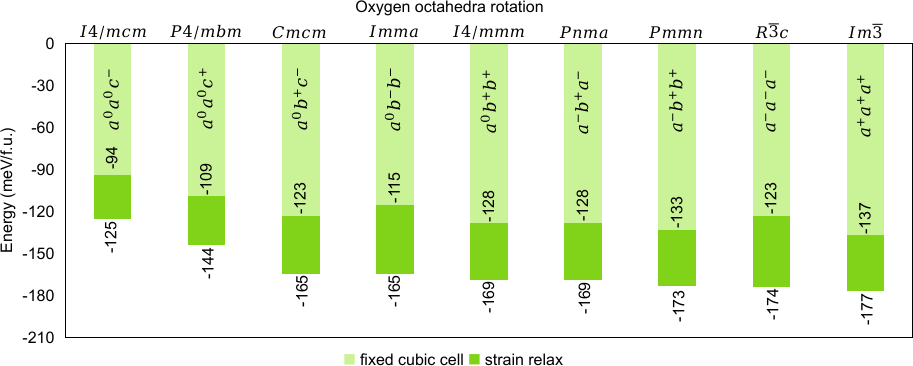}} \\
\multicolumn{2}{c}{{\small (b)}}   
\end{tabular}
\caption{{\small (Color online) Energy (meV/f.u.) of different relaxed phases of AgNbO$_3$, respect to the equilibrium cubic paraelectric {\it Pm$\bar{3}$m} state taken as reference.  Atomic relaxations were performed using the PBEsol functional keeping the original cubic cell fixed (light colors) or fully relaxing the unit cell (dark color). The different phases are characterized using generalized Glazer's notations with, along each direction, the superscript labeling the kind of oxygen-octahedra rotation ($a^+$ for in-phase $M_3^+$ rotations, $a^-$ for anti-phase $R_4^+$ rotation and $a^*$ for modulated $T_4$ rotations) and the $p$ subscript indicating the presence of a polarization. 
}}
\label{fig:figure3}
\end{figure*}
\section{Structural Instabilities}\label{sec:allinstability}
We now turn our attention to the condensation of the unstable modes identified in the phonon spectra of the high-symmetry cubic phase. Our goal is to understand how these modes behave both individually and in combination, and to assess the role of mode coupling in stabilizing specific low-temperature phases.
We organize the discussion by first discussing polar modes, then oxygen-octahedra rotations, and finally their combinations, highlighting how each contributes to the stabilization of the  AgNbO$_3$ ground state.

\subsection{Polar and antipolar instabilities} 
Looking at the phonon dispersion curves of the cubic phase (Fig.~\ref{fig:figure1}), the most unstable mode is the polar $\Gamma_4^-$ mode (175{\it i} cm$^{-1}$).
Consistent with the large anomalous Born effective charges discussed above, the eigendisplacements of the $\Gamma_4^-$ mode reveal a predominantly B-site–driven polar distortion, typically associated to compounds with a tolerance factor larger or close to 1. In this mode, predominantly Nb atoms move out of phase with respect to the oxygen sublattice, confirming that the instability originates primarily from Nb–O hybridizations. The associated mode-effective charge amounts to 8.50$e$, exceeding the value reported for NaNbO$_3$ (7.88$e$) and highlighting the strong polar character of AgNbO$_3$.

Figure~\ref{fig:figure3}(a) shows the calculated polarization and relative energies of AgNbO$_3$ when only the polar instability is condensed. The {\it P$4$mm}, {\it Amm$2$}, and {\it R$3$m} structures correspond to the condensation of the unstable ferroelectric mode along the [100], [110], and [111] directions, respectively.
If only the atomic positions are relaxed while keeping the cubic cell fixed, the usual ferroelectric hierarchy is recovered: the tetragonal phase is highest in energy, followed by the orthorhombic phase, with the rhombohedral phase being the most favorable. However, once full structural relaxation is allowed i.e. including both internal coordinates and strain, the energetic sequence is modified by polarization–strain coupling. In this case, the orthorhombic {\it Amm$2$} phase becomes the lowest-energy ferroelectric structure within GGA-PBEsol (Fig.~\ref{fig:figure3}(a)), similar to the situation in NaNbO$_3$.
The fully relaxed {\it Amm$2$} phase exhibits a spontaneous polarization of 48.8~$\mu$C/cm$^2$ (to be compared with the LDA value of 55.9~$\mu$C/cm$^2$ reported in Ref.~\cite{niranast}), comparable to the experimental high-field polarization of 52~$\mu$C/cm$^2$ measured in AgNbO$_3$ polycrystals~\cite{futaito}. 

Turning to the condensation of individual anti-polar modes associated to the unstable branch extending from the ferroelectric instability, the gains of energy are relatively similar, although always smaller than for the polar mode : $\Delta_5$ ($-$22.4 meV/f.u.), $X_5^+$ ($-$15.4 meV/f.u.), $M_3^-$ ($-$15.9 meV/f.u.). Overall, these gains of energy appear relatively modest in view of the significant amplitude of the instabilities ($\approx 200 i$ cm$^{-1}$) and highlight a strongly anharmonic energy landscape. These gains of energy are however comparable to those reported in other B-type ferroelectrics like BaTiO$_3$ or KNbO$_3$.

\subsection{Antiferrodistortive instabilities} 
We now switch to a detailed analysis of the instabilities of the antiferrodistortive branch extending from $R$ to $M$ and their role in shaping the low-energy phase diagram.
It is well established that such modes related to rotation of the oxygen octahedra are among the most ubiquitous structural instabilities in perovskites~\cite{glazer1,glazer2,woodward,howardst}. In Glazer notation, these distortions at the Brillouin zone boundary are associated to in-phase and anti-phase rotations of the oxygen octahedra about the pseudocubic axes. In our case, in-phase rotations transform like the $M_3^+$ irreducible representation ($q = \frac{1}{2},\frac{1}{2},0$), while anti-phase rotations correspond to $R_4^+$ ($q = \frac{1}{2},\frac{1}{2},\frac{1}{2}$)~\cite{howardst}.

In cubic AgNbO$_3$, both in-phase and anti-phase oxygen-octahedra rotations are unstable, with significant imaginary frequencies ($M_3^+$: 162{\it i} cm$^{-1}$; $R_4^+$: 156{\it i} cm$^{-1}$). This suggests already that octahedra rotational distortions are expected to play a central role in determining the ground state.
To assess their intrinsic energetic hierarchy, we systematically condense the $M_3^+$ and $R_4^+$ modes independently and in combination along all symmetry-allowed directions, generating the full set of possible tilt patterns: $(a^0a^0c^+)$, $(a^0a^0c^-)$, $(a^0b^+b^+)$, $(a^0b^-b^-)$, $(a^0b^+c^-)$, $(a^+a^+a^+)$, $(a^-b^+b^+)$, $(a^-b^+a^-)$, and $(a^-a^-a^-)$ in Glazer's notations. The corresponding fully relaxed energies are shown in Fig.~\ref{fig:figure3}(b).

Two main conclusions emerge. First, the condensation of in-phase $M_3^+$ rotations (which appear also to be slightly more unstable than anti-phase one in Table~\ref{tab:table2}) are systematically producing larger decrease of energy than antiphase $R_4^+$ rotations, whether condensed along one, two, or three crystallographic directions. 
Second, combinations containing a larger in-phase component are systematically lower in energy than those dominated by antiphase rotations. For instance, the $a^-b^+b^+$ ({\it Pmmn}) structure is more stable than the $a^-b^+a^-$ ({\it Pnma}) configuration. Even in the absence of strain relaxation, oxygen-octahedra rotation patterns with dominant in-phase character remain energetically preferred over purely antiphase configurations such as $a^-a^-a^-$ ({\it R$\bar{3}$c}). 
Consistently with that, in Fig.~\ref{fig:figure3}(b), the $a^+a^+a^+$ ($Im\bar{3}$) rotation pattern appears as the one that produces the largest lowering of energy. This trend is particularly remarkable and at odds with what is usually observed in other perovskites in which anti-phase rotations typically dominate~\cite{Benedek2013} .

Interestingly, although the polar instability is stronger than the antiferodistortive ones in the cubic phase (Fig.~\ref{fig:figure1}), the energy gains associated with oxygen-octahedra rotations (ranging from roughly 125 to 177 meV/f.u., depending on the pattern) are substantially larger than those obtained for the purely ferroelectric $\Gamma_4^-$ mode ($\sim$31 meV/f.u.). This confirms  that, as previously reported~\cite{Huazhang-24}, in AgNbO$_3$, oxygen-octahedra rotation is the dominant structural driving force for decreasing the energy, while polar/antipolar distortions are more anharmonic and provide more modest stabilization.

\subsection{Mode coupling and low-energy phases}\label{sec:grstatinstabilit}
Having analyzed the condensation of polar and antiferrodistortive instabilities separately, we now turn to their combined effect. In AgNbO$_3$, the stabilization of the lowest-energy phases cannot be understood in terms of isolated distortions, but rather through the anharmonic coupling between polar, antipolar, and antiferrodistortive modes.
In the following, we examine the fully relaxed structures obtained by simultaneously condensing the relevant AFD and polar or antipolar modes, and analyze how their interplay selects the true low-energy configurations.

\begin{figure}[htbp!]
\centering
\includegraphics[width=8cm,keepaspectratio=true]{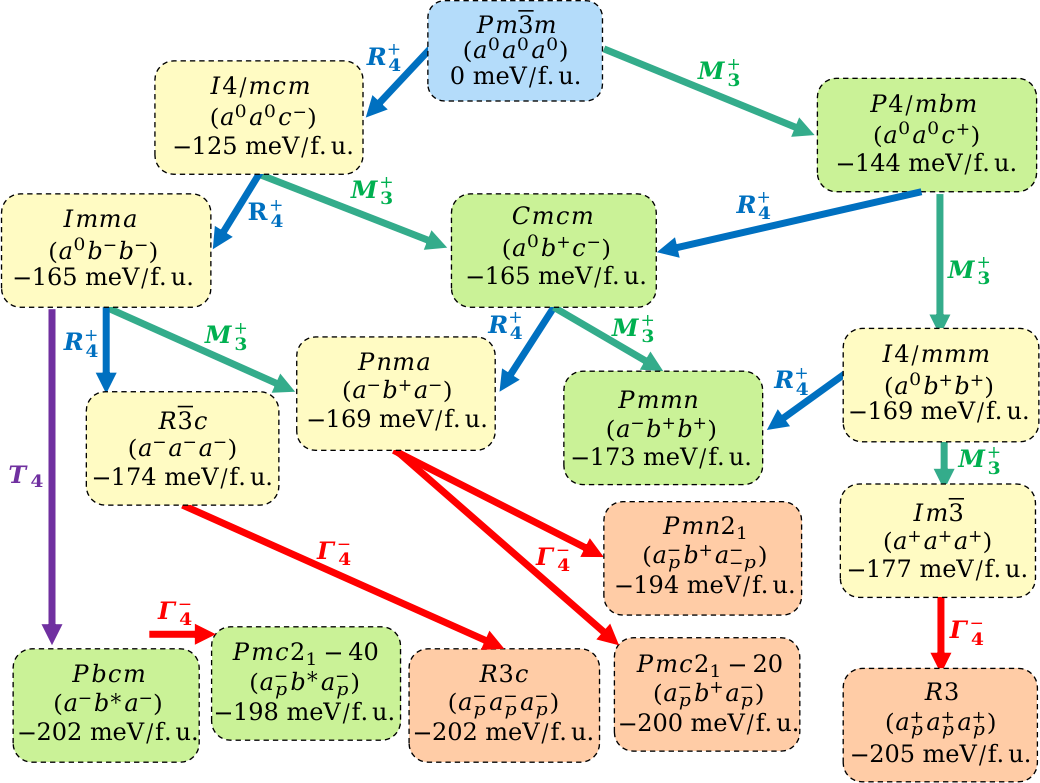}
\caption{{\small (Color online) Comparison of the energies (meV/f.u.) of various metastable phases of AgNbO$_3$, respect to the cubic structure taken as reference (using the PBEsol functional).The green boxes labels the phases experimentally observed. The orange boxes highlights additional competitive low-energy polar phases, including the $R3$ ground state. The different phases are characterized using generalized Glazer's notations with, along each direction, the superscript labeling the kind of oxygen-octahedra rotation ($a^+$ for in-phase $M_3^+$ rotations, $a^-$ for anti-phase $R_4^+$ rotation and $a^*$ for modulated $T_4$ rotations) and the $p$ subscript indicating the presence of a polarization along a given direction. }}
\label{fig:figure5}
\end{figure}

Figure~\ref{fig:figure5} provides a schematic overview of the stabilization pathways of AgNbO$_3$ from the cubic reference. The diagram illustrates how different combinations of polar, antipolar, and rotational modes condense to produce various  low-energy phases of  {\it Pmn$2_1$},{\it Pmc$2_1$},  {\it R$3$c}, {\it Pbcm}, and  {\it R$3$} symmetries.

\begin{figure}[htbp!]
\centering
\begin{tabular}{cc}
 {\small (a)} &\includegraphics[width=7cm,keepaspectratio=true]{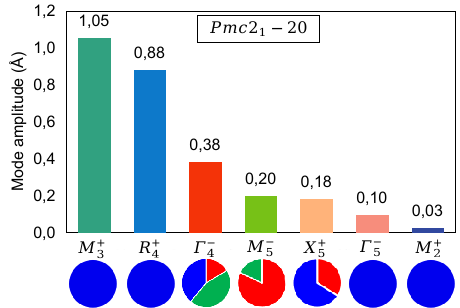}\\ {\small (b)} & \includegraphics[width=7cm,keepaspectratio=true]{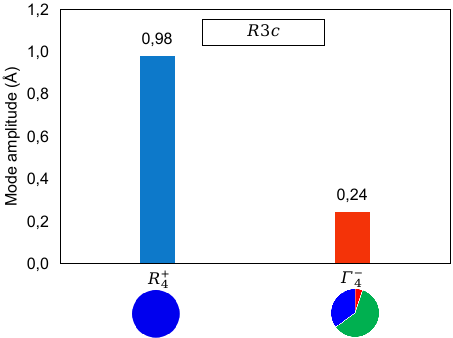} \\
{\small (c)} & \includegraphics[width=7cm,keepaspectratio=true]{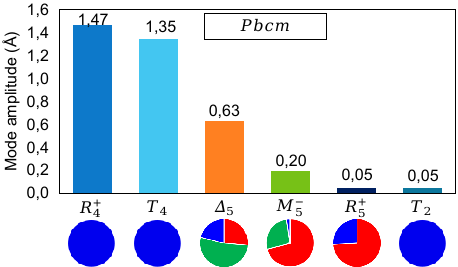} \\
{\small (d)} & \includegraphics[width=7cm,keepaspectratio=true]{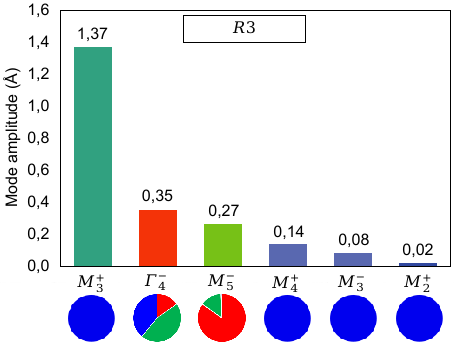}  
\end{tabular}
\caption{{\small (Color online) Contributions  of symmetry-adapted modes [\AA] to the atomic distortion of different metastable phases respect to the cubic phase (main bar charts) and related atomic contributions to each mode (bottom pie charts, with Ag in red, Nb in green and O in blue), for the {\it Pmc$2_1-20$}, {\it R$3$c}, {\it Pbcm}, and {\it R$3$}  phases of AgNbO$_3$. Results obtained using the PBEsol functional.
}}
\label{fig:figure4}
\end{figure}

Guided by this roadmap (Fig.~\ref{fig:figure5}), we examine these states in detail. Our GGA-PBEsol calculations indicate that distinct structures -- {\it Pmc$2_1$}, {\it R$3$c},  {\it Pbcm}, and {\it R$3$} -- converge at remarkably close energies, within approximately 5 meV per formula unit of each other. In particular and further discussed later, the fully relaxed {\it Pbcm} phase is lower in energy than {\it Pmc$2_1$}-40 (by 4.0 meV/f.u.)  and {\it R$3$c} (by 0.6 meV/f.u.), in good agreement with the findings of Moriwake {\it et al.}~\cite{Moriwake}. Minor discrepancies with other theoretical studies~\cite{{niranast},{zhangzhi},{shigwada}}, which report slightly different energetic orderings, can be attributed to small variations in computational details. 
Nevertheless, among all candidates, the previously overlooked rhombohedral {\it R$3$} structure emerges as the unequivocal thermodynamical ground state  of AgNbO$_3$, being located 2.1 meV/f.u. below the $Pbcm$ phase (i.e. $\Delta E_{RP} = -2.1$ meV/f.u.). This result appears to be robust. Although tiny, we checked carefully that $\Delta E_{RP}$ is fully converged in terms of energy cutoff and $k$-point sampling. Moreover, the result remains unchanged  using the LDA ($\Delta E_{RP} = -0.8$ meV/f.u., see Appendix B) or GGA-PBE ($\Delta E_{RP} = -2.6$ meV/f.u.)  energy functionals that have the tendency respectively to underestimate and overestimate the lattice parameters. As such, and although the energies of $R3$ and $Pbcm$ are almost degenerated, our result appears robust at the DFT level. Moreover, we checked that the $R3$ phase does not show any phonon instability and is dynamically stable, which is however also the case of the $Pbcm$ phase (see Appendix~\ref{sec:dynstability}).

To characterize these phases in detail, we performed a mode-by-mode decomposition of their total distortion (Fig.~\ref{fig:figure4}) and analyze the atomic contributions of the individual modes. In Appendix C, we report the lattice parameters and distortion amplitudes of the lowest-energy phases, providing a reference for both experimental and computational comparisons.

As usual in perovskites showing AFD instabilities ($t<1$), the $Pnma$ phase ($a^-b^+a^-$) appears at low energy. Then, following trends reported by Benedek and Fennie when $t \approx 1$~\cite{Benedek2013}, the $R\bar{3}c$ phase ($a^-a^-a^-$) is lower in energy. Surprisingly, here, the $\Gamma_4^-$ ferroelectric distortions that is expected to compete with AFD, remain unstable in the latter, yielding a ferroelectric  {\it R$3$c} phase ($a^-_pa^-_pa^-_p$) significantly lower in energy and showing a macroscopic polarization of 44.8 $\mu$C/cm$^2$ (51.1 $\mu$C/cm$^2$ \cite{polarization_note}). 
This can eventually be related to the B-type character of the ferroelectric distortion, which competes less with AFD than A-type motions. 

Starting from the {\it Pnma} phase ($a^-b^+a^-$) and condensing a polar distortion in the plane of both anti-phase rotations, give rise to metastable polar phases of $Pmn2_1$($a^-_pb^+a_{-p}^-$) and $Pmc2_1$ ($a^-_pb^+a^-_{p}$) symmetries, with orthogonal polarization and 20 atoms per unit cell. We so refer to the latter as $Pmc2_1-20$ since another $Pmc2_1$ phase will be discussed later. It shows a spontaneous polarization of 54 $\mu$C/cm$^2$~\cite{polarization_note}. Both these phases are further stabilized by the activation of anti-polar secondary modes (Fig. \ref{fig:figure4}) but remain higher in energy than $R3c$.  

Starting now from the {\it Imma} phase ($a^0b^-b^-)$ and condensing instead along the third direction  the $T_4$ rotation pattern ($++--$, further labeled $b^*$) produces the non-polar {\it Pbcm} phase ($a^-b^*a^-$) typically observed at room temperature. The stabilization of this {\it Pbcm} phase at an energy lower than $R3c$ has previously been rationalized from anharmonic mode couplings with antipolar $\Delta_5$ distortions~\cite{Huazhang-24}.

Condensing polar distortions in the $Pbcm$ phase drives the system to another polar $Pmc2_1$ phase ($a^-_pb^*a^-_p$) with 40 atoms per cell,  labeled $Pmc2_1-40$. This phase has been previously computed~\cite{Moriwake} and has also been experimentally observed~\cite{{tzhang},{yashima},{changszyf}}. It shows a polarization of 49 $\mu$C/cm$^2$ (51.5 $\mu$C/cm$^2$~\cite{polarization_note}), close to the experimental value of 52 $\mu$C/cm$^2$. This phase appears few meV/f.u. higher in energy than $Pbcm$ but it has been suggested that both $Pbcm$ and $Pmc2_1-40$ can coexist at room temperature, explaining the weak ferroelectric character of AgNbO$_3$. Still the non-polar $Pbcm$ phase is usually considered as the thermodynamical ground state of AgNbO$_3$, while the proximity of the $Pmc2_1-40$ 
(and $R3c$) polar phases contributes to explain its antiferroelectric character.

However, starting instead from the $Im{\bar 3}$ phase with the most favorable $a^+a^+a^+$ oxygen rotation pattern (Fig. \ref{fig:figure3}b) and condensing a polar distortion along the $[111]$ direction, we reach a polar {\it R$3$} phase ($a^+_pa^+_pa^+_p$), located at an energy not only lower than {\it R$3$c} but also lower than the expected {\it Pbcm} ground state. This highlights that, in amazing resonance with  PbZrO$_3$~\cite{Aramberri2021}, the correct thermodynamical ground state of antiferroelectric AgNbO$_3$ might in fact be ferroelectric!
This conclusion does not necessarily contradict the fact that the $R3$ phase has remained experimentally elusive thus far. Indeed, the $Pbcm$ structure lies only a few meV above the $R3$ ground state and is itself dynamically stable. Such a near-degeneracy suggests that kinetic effects may play an important role in determining the experimentally realized phase. In particular, if the $Pbcm$ phase condenses first upon cooling, the system may become trapped in this local minimum because of the energy barrier separating $Pbcm$ and $R3$, thereby preventing the transformation toward the true thermodynamic ground state. A similar situation has been proposed for PbZrO$_3$~\cite{Huazhang-24b}, where the low-temperature $Ima2$ ferrielectric phase is energetically competitive with $Pbam$, but its realization is kinetically hindered by the energy barriers separating these states, resulting in the persistence of the $Pbam$ structure down to low temperature. In this perspective, the absence of an experimental observation of the $R3$ phase in AgNbO$_3$ may reflect a similar kinetic trapping.

The $R3$ phase of AgNbO$_3$ combines significant in-phase rotations of the oxygen octahedra (7.7$^o$) and a large polarization of 44.5 $\mu$C/cm$^2$ (49.0 $\mu$C/cm$^2$~\cite{polarization_note}). Different factors can be evoked to rationalize the appearance of this unusual phase : 
at first, the uncommon predominance of in-phase rather than out-of phase oxygen octahedra rotation; then, the B-type character of the ferroelectric instability that contributes to reduce its competition with coexisting AFD;  finally, the appearance of distinct secondary modes, which further decrease the energy. In that respect, as illustrated in Fig. 4d, the $R3$ phase condenses a substantial amplitude of $M_5^-$ anti-polar motions. However, the $M_5^-$ mode is stable in the $Pm\bar{3}m$ structure so that it should not naturally appear. Its condensation in the $R3$ phase is driven by primary $M_3^+$ in-phase rotations and $\Gamma_4^-$ polar motions through a trilinear energy term of the form $E \propto \lambda Q_{M_3^+}Q_{\Gamma_4^-}Q_{M_5^-}$, evoking what appears in some improper ferroelectrics~\cite{Bousquet-08}.   

\section{Emergence of chirality in the R3 phase}

An intriguing feature of the hidden rhombohedral $R3$ ground state of AgNbO$_3$ is the emergence of structural chirality, a property absent in its other low-energy phases.
The trigonal space group $R3$ is one of the 65 Sohncke space groups, which are defined by the presence of only proper symmetry operations (rotations and translations) and the absence of improper operations such as inversion, mirror planes, or rotoinversion axes.

Because these space groups contain only orientation preserving symmetry operations they are compatible with chirality. 
Interestingly, the crystals with the $R3$ space group can be chiral but $R3$ is a non-enantiomorphic Sohncke space group: unlike enantiomorphic pairs (such as $P3_1$ and $P3_2$), $R3$ can accommodate both left and right handed structures~\cite{fecher2022, Bousquet-25}. 
As a result, crystallization in the $R3$ space group does not inherently guarantee a chiral structure, rather, it preserves the handedness of any chiral motif present.
In other words, an achiral motif can still yield an achiral crystal structure in this space group.

\begin{figure}[h]
     \centering
      \includegraphics[width=\columnwidth]{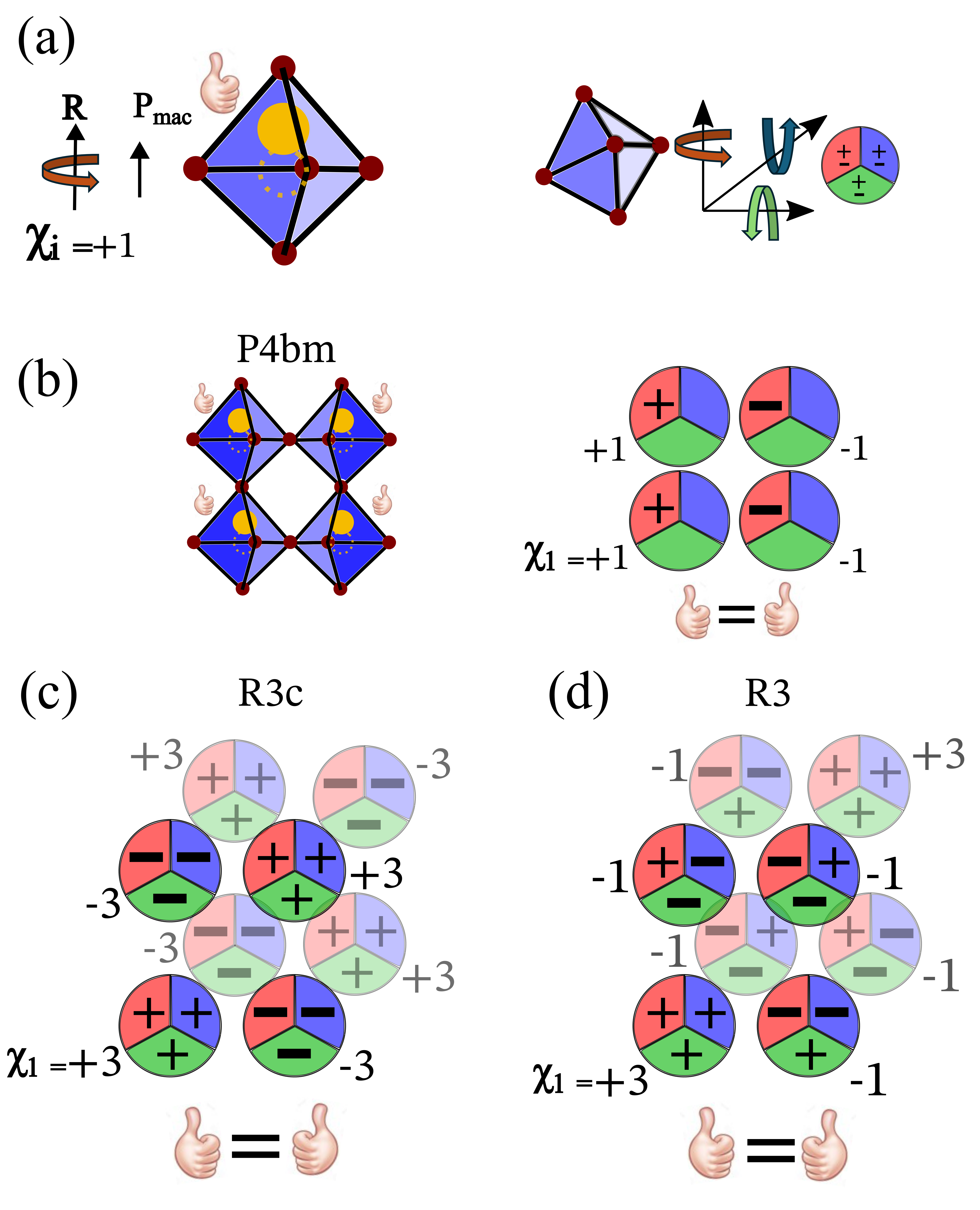}
      \caption{Schematic representation of the chiral distribution of the structural motif in different phases of AgNbO$_3$. (a) Left: Octahedral rotations $\mathbf{R}$ combined with a polar displacement $\mathbf{u}$ along the rotation axis confer a local intrinsic chirality to each perovskite unit. Right: Mapping of a rotation pattern onto a circle divided in three sectors. The green, blue, and red colors refer to rotations along the $x$, $y$, and $z$ axes, respectively, while the $+$ and $-$ symbols on each sector indicate the presence of in-phase and out-of-phase rotations along the corresponding direction. (b) $P4bm$ phase $(a^0,a^0,c^+_p)$, characterized by $(a^0,a^0,c^+)$ octahedral rotations and polarization along $[001]$, resulting in an overall achiral structure due to pairwise cancellation of local chirality $\pm 1$. (c) $R3c$ phase $(a^-_p,a^-_p,a^-_p)$, characterized by $(a^-,a^-,a^-)$ rotations and polarization along $[111]$, also yielding a globally achiral structure through pairwise cancellation of chirality $\pm 3$. (d) $R3$ phase $(a^+_p,a^+_p,a^+_p)$, characterized by $(a^+,a^+,a^+)$ rotations and polar displacements along $[111]$. Within the purely geometric approximation, the local chiral motifs cancel out, yielding no net chirality.} 
      \label{fig:chir} 
\end{figure}

In AgNbO$_3$, the $R3$ phase ($a^+_pa^+_pa^+_p$) results from the condensation of oxygen octahedra rotations $a^+a^+a^+$ and polarization along the $[111]$ direction. 
In such a case, as illustrated in Fig.~\ref{fig:chir}(a), each {\it individual} perovskite unit $i$ undergoes a rotation, $\mathbf{R}_i$, combined with a polar displacement $\mathbf{u}_i$ along the same axis granting the system with a macroscopic polarization $\mathbf{P}_{\textrm{mac}}$ and forming a local motif in each unit cell $i$ that is intrinsically chiral, with non-vanishing $\chi_i = \mathbf{R}_i \cdot\mathbf{P}_i$: the axial nature of the rotation and the polar nature of the displacement prevent the motif from being superimposed on its mirror image by any proper rotation~\cite{hlinka2014}. 

However, a local chiral motif, $\chi_i$, does not necessarily result in a globally chiral structure with non-vanishing $\chi=\sum_i \chi_i$. 
This is obvious, for example, in a fictitious $P4bm$ or the $R3c$ phase.
A $P4bm$ phase $(a^0a^0c^+_p)$  resulting from a combination of $(a^0a^0c^+)$ rotation pattern and $[001]$ polar distortions, with axial and polar axes aligned along the $z$ direction, would exhibit  a local non-zero chirality (Fig. \ref{fig:chir}b)~\cite{Gomez-Ortiz-24}.
However, in neighboring cells along the $x$ and $y$ directions, the octahedra rotate in the opposite direction, reversing the axial vector and thus the chirality, while the polar vector remains unchanged. Consequently, the opposite local chiralities of neighboring cells cancel out two by two and the structure is globally achiral.
In the physically more relevant $R3c$ phase $(a^-_pa^-_pa^-_p)$, combining $(a^-a^-a^-)$ rotation and polarization in the  $[111]$ direction, the alternating oxygen octahedra rotations in neighboring cells also leads to the cancellation of local chiralities, resulting in a net achiral or anti-chiral structure. 

\begin{figure}[h]
     \centering
     \includegraphics[width=\columnwidth]{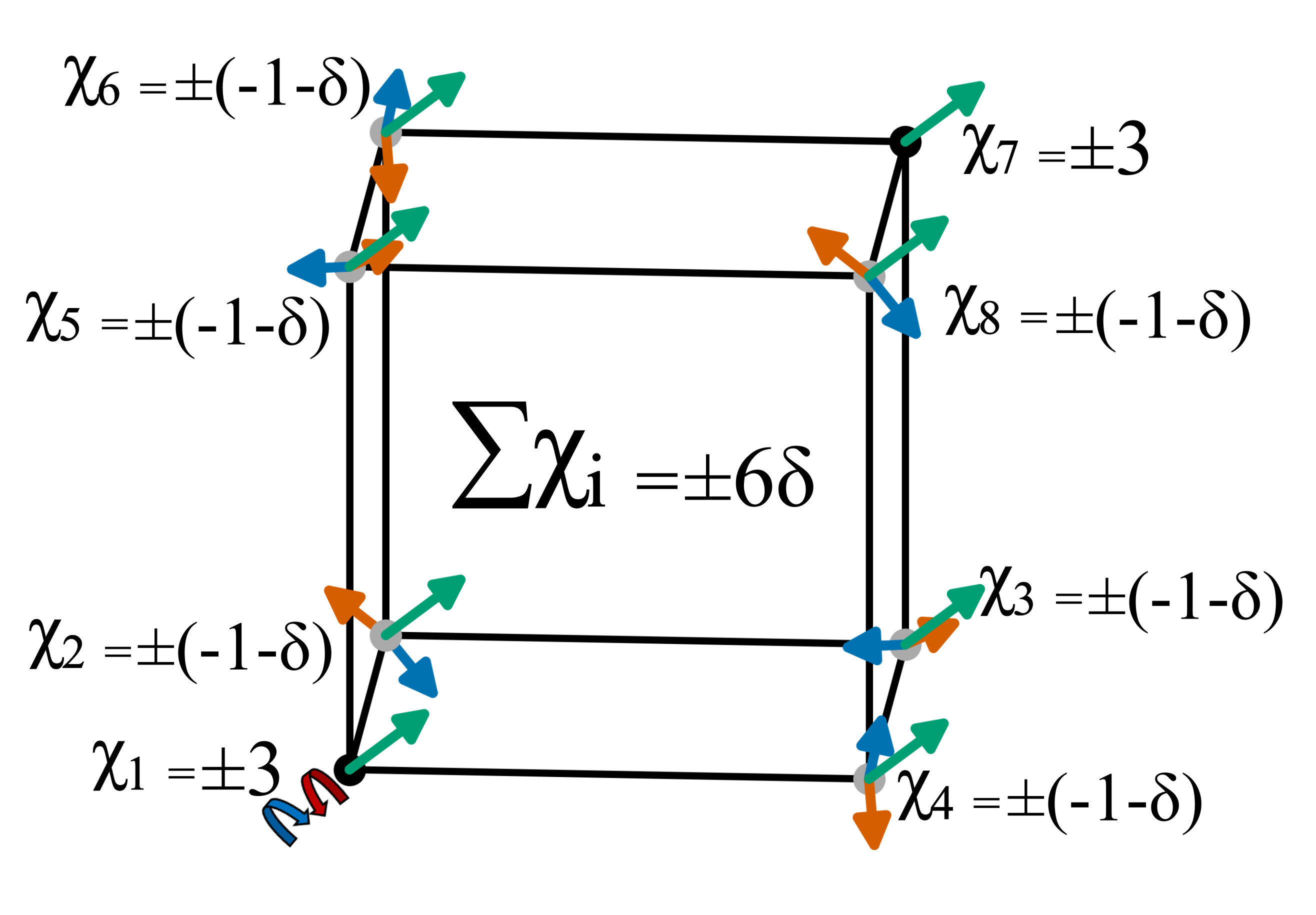}
      \caption{Local polarization directions in an artificial $R3$ structure condensing only the polar disortions along the [111] direction and the ($a^+a^+a^+$) oxygen octahedra rotations. Green arrows denote the local polarization vectors obtained using the cubic-phase Born effective charges, for which all local polarizations are aligned with the $[111]$ direction. Blue and orange arrows show the transverse components of the local polarizations, perpendicular to the former cubic-phase polarization direction, obtained using the actual Born effective charges of the two enantiomorphic $R3$ variants (for clarity, these deviations have been magnified by a factor of five). The two variants are related by reversal of the octahedral rotational mode while keeping the polar mode fixed. These local deviations break the exact cancellation of neighboring chiral contributions predicted by the purely geometric picture Fig.~\ref{fig:chir}, resulting in a finite net chirality of opposite sign ($\pm$) for the two enantiomorphs. The global picture remains almost identical in the fully relaxed $R3$ phase additionally condensing $M_5^-$ antipolar motions.} 
      \label{fig:poldeviate} 
\end{figure}

From a general perspective, since octahedra rotations are antiferrodistortive, the direction of rotation alternates between neighboring units.
When combined with a polar distortion, this alternation creates neighboring motifs with opposite handedness, forming an \textit{antichiral} pattern with no net chirality.
However, if an octahedra rotation pattern is combined with antipolar motions, a non-zero net chirality can emerge, as both ``anti-'' components align in the same direction from site to site, summing to produce chirality rather than canceling it~\cite{Bousquet-25,Hayashida-21}.
According to this analysis, one might initially expect the same cancellation of chirality in the $R3$ phase of AgNbO$_3$ ($a^+a^+a^+$ rotation combined with $[111]$ polar displacement).
Indeed, as illustrated in Fig.~\ref{fig:chir}(d), computing  {\emph{local}} chiralities $\chi_i = \mathbf{R}_i \cdot\mathbf{P}_{\rm mac}$ from the coupling between the three in-phase rotations $\mathbf{R}_i $ with a uniform macroscopic diagonal polarization $\mathbf{P}_{\rm mac}$ results in the erroneous conclusion of a net cancellation of chirality.
To understand why this phase exhibits net chirality and non-zero optical activity, one should better compute the {\emph{local}} chiralities $\chi_i$ using the {\emph{local}} polarization at each perovskite unit as $\mathbf{P}_{\rm{loc},i} = (1/\Omega_0) \sum_{\kappa} Z^*_{\kappa,i} \mathbf{u}_{\kappa,i}$ where the sum runs over the atoms of each cell and $Z^*_{\kappa}$ are the Born effective charges. As illustrated in Fig.~\ref{fig:poldeviate}, when the local polarizations at each unit cell are properly computed with the actual Born effective charges of the $R3$ phase, the local polarizations $\mathbf{P}_{\rm{loc},i}$ are no longer strictly aligned with the [111] direction (although their sum still yields a [111] macroscopic polarization). As a result of these deviations, the chiral contributions from neighboring unit cells no longer cancel out exactly (Fig.~\ref{fig:poldeviate}), ultimately giving rise to a \emph{ferrichiral} state. Interestingly, using instead the diagonal Born effective charges of the cubic phase (green arrows in Fig.~\ref{fig:poldeviate}) yields the same local $\mathbf{P}_{\rm{loc},i}$ at all sites and a net cancellation of chirality. Consequently, the emergence of chirality in the $R3$ phase can be seen as a purely electronic effect that cannot be anticipated from the inspection of the atomic distortions : it arises from the non-diagonal component of the Born effective charges, reflecting the anisotropic rearrangement of the electrons under the atomic distortions. 
This electronic origin of the chirality is further reinforced by the fact that, as illustrated in Fig.~\ref{fig:poldeviate}, the previous discussion already applies to an artificial $R3$ phase condensing only non-chiral polar distortions and oxygen octahedra rotations. As further discussed below, the fully relaxed $R3$ phase further condenses a chiral anti-polar $M_5^-$ distortion; the latter further tilts the local polarization from the [111] directions but at the end only marginally affects the net chirality and natural optical activity (i.e. it amplify it by only 15\%).  We notice that this phenomenon is specific to the $R3$ phase and not observed in the achiral $P4bm$ or $R3c$ phases previously discussed.

According to its chiral character, the $R3$ phase of AgNbO$_3$ exhibits significant natural optical activity (NOA). 
For light propagating along the $C_3$ axis, we find a value of $g_{33}=2.19$ deg/(mmeV$^2$) as computed with {\sc{abinit}}~\cite{{Zabalo-23}}, which is consistent with the emergence of a ferrichiral state. 
This value is strikingly comparable to that observed in quartz~\cite{{Jonson-96},{Zabalo-23},{Gomez-Ortiz-26}}, a prototypical material renowned for its NOA. 
It is important to note that the physically relevant quantity for comparison with experiments is the symmetrized $g_{ij}$ tensor that is diagonal in the $R3$ space group, and whose component along the principal axis corresponds to $g_{33}$. 
Interestingly, if either the oxygen octahedra rotation or polarization pattern is individually reversed, this component reverses its sign, further confirming the emergence of net chirality in the $R3$ phase as the bi-product of these two quantities. This is further illustrated in Fig.~\ref{fig:poldeviate}, which higlights how local polarizations are affected under reversal of the oxygen octahedra rotations.

It should be noted that semilocal DFT functionals such as PBEsol are known to underestimate electronic band gaps, which may affect calculated optical response functions. To assess the sensitivity of the present results, we applied a scissor correction within {\sc{abinit}}. For $\alpha$-quartz, a correction reproducing the experimental band gap decreases the calculated NOA by about 30\%{}, resulting in slightly worse agreement with experiment than the uncorrected calculation. This suggests that, although a scissor correction improves the electronic band gap, it does not necessarily improve the agreement between calculated and experimental optical activity, as discussed in the literature~\cite{Zabalo-23}. For the $R3$ phase of AgNbO$_3$, the experimental band gap is currently unknown. Nevertheless, using representative scissor corrections of 1.00 and 1.24 eV, corresponding to the corrections required to reproduce the experimental band gaps of the cubic and Pbcm phases, decreases the calculated NOA by approximately 35\%{} and 40\%{}, respectively. Thus, although the absolute value is affected by the band-gap correction, the predicted NOA remains comparable to that of quartz.

Finally, it is worth noticing that the emergence of chirality in AgNbO$_3$ can be attributed to an improper mechanism. $M_5^-$ antipolar motions of the cubic $Pm\bar{3}m$ reference phase, appearing in the $R3$ phase (Fig. \ref{fig:figure4}), are chiral in the sense that their condensation along the $(a,b;a,b;a,b)$ direction will bring the system in the $R3$ chiral state. 
However, this $M_5^-$ mode is not unstable (Fig. \ref{fig:figure1}) and cannot spontaneously condense to produce the ground state. 
Instead, as previously discussed, the phase transition arises from the joint condensation of the polar ($\Gamma_4^-$) and antiferrodistortive ($M_3^+$) unstable modes. 
When considered individually, these modes are achiral, but when considered together along the [111] direction they confer chirality to the system. 
In other words, the product of their irreducible representation is $M_5^-$ ($\Gamma_4^- \otimes M_3^+ \equiv M_5^-$) and the joint appearance of these two achiral soft modes will naturally drive the additional hard chiral $M_5^-$ motions from the trilinear coupling term previously discussed ($E \propto \lambda Q_{M_3^+}Q_{\Gamma_4^-}Q_{M_5^-}$).
This allows us to make a direct parallelism with a specific type of improper ferroelectricity where two non-polar soft modes can induce polarization to the system from their trilinear coupling with a hard polar mode~\cite{Bousquet-08}. 
We notice however that, in contrast with what is often observed in such improper ferroelectrics, where the polarization mostly arises from the condensation of the driven polar mode, in the present case, the system already shows a significant NOA when condensing the polar and AFD modes only while  the NOA is only marginally affected by the additional appearance of the chiral anti-polar motions.

\section{Conclusions}\label{sec:conclusion}
In summary, we have shown that the rhombohedral $R3$ ferroelectric phase is the proper thermodynamic ground state of AgNbO$_3$. Although the energy difference with the conventionally accepted antiferroelectric $Pbcm$ ground state is very small, making the two phases nearly degenerate in practice, this energetic ordering is robust across different exchange–correlation functionals (LDA, PBEsol, and PBE), despite their well-known tendency to underestimate, accurately reproduce, or overestimate the equilibrium volume, respectively. The two phases therefore likely compete at low temperature.
This $R3$ phase arises from the cooperative condensation of a polar distortion and an unusual pattern of in-phase rotations of the oxygen octahedra, both aligned along the $[111]$ direction, which is energetically favored over the more common out-of-phase rotations.
Such behavior reflects a general trend in antiferroelectric perovskites, where flat energy landscapes often give rise to competing ferroelectric phases or even ferroelectric ground states.

Remarkably, although one might expect that the antiferroaxial nature of octahedral rotations combined with a polar mode would cancel local chiralities, the in-phase rotations along all three axes in the $R3$ phase produce an incomplete cancellation, giving rise to a ferrichiral structure.
This mechanism naturally explains the emergence of significant natural optical activity in the $R3$ phase, and provides a clear microscopic picture of how the main polar and in-phase rotational distortions give rise to chirality. Moreover, it also explains its improper nature: chirality does not arise from an intrinsically unstable chiral mode, but from the cooperative coupling of the achiral polar and in-phase rotational distortions, which naturally induces secondary chiral anti-polar motions driven by the stable $M_5^ -$ mode along the $(a,b;a,b;a,b)$ direction. In this sense, the emergence of chirality is analogous to improper ferroelectricity, where the combination of non-polar instabilities gives rise to a new macroscopic property.

Overall, this work illustrates how subtle couplings between competing structural instabilities can give rise to emergent phenomena in antiferroelectrics, motivating experimental exploration of AgNbO$_3$ and related compounds.
\section*{Acknowledgments}
This work has been supported by the Erasmus$+$International Credit Mobility program 2023 of Universit\'e{} de Li\`e{}ge and the PRD-CCD ARES 2019-2024 project: {\it Le coltan du Kivu: Capacit\'e{} de traitement physico-chimique et \'e{}tudes d'applications}. 
F.G.O. acknowledges financial support from MSCA-PF 101148906 funded by the European Union and the Fonds de la Recherche Scientifique (FNRS) through the FNRS-CR 1.B.227.25F grant. 
PhG acknowledges financial support from the F.R.S.-FNRS Belgium through the PDR project ToPoTex (grant No T.0128.25).
We acknowledge the use of the CECI supercomputer facilities funded by the F.R.S-FNRS (Grant No. 2.5020.1) and of the Tier-1 supercomputer of the Fédération Wallonie-Bruxelles funded by the Walloon Region (Grant No. 1117545).
\clearpage
\newpage
\appendix
\setcounter{figure}{0}
\setcounter{table}{0}
\renewcommand{\thefigure}{A\arabic{figure}}
\renewcommand{\thetable}{A\arabic{table}}
\section{Phonon mode labels}
\label{sec:phononlabels}
\begin{figure}[htbp!]
\centering
\begin{tabular}{cc}
 \includegraphics[width=3.0cm,keepaspectratio=true]{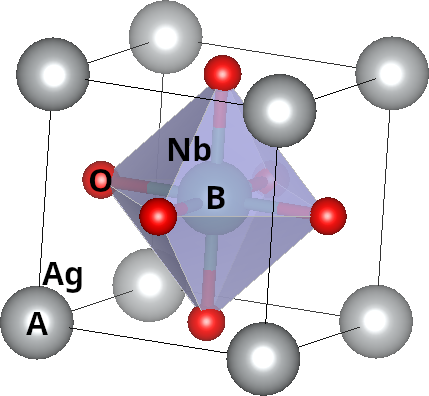} & \includegraphics[width=3.0cm,keepaspectratio=true]{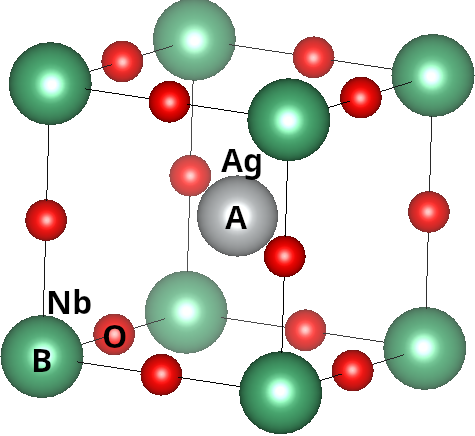}\\
 {\small (a)} & {\small (b)}  
\end{tabular}
\caption{{\small (Color online) Two different ways of representing the cubic {\it Pm$\bar{3}$m} perovskite structure: (a) {\it A} at 0.0 0.0 0.0 (1a), vertices of cubes, {\it B} at 0.5 0.5 0.5 (1b), centers of cubes, and {\it O} at 0.5 0.5 0.0 (3c), faces of cubes. (b) {\it A} 0.5 0.5 0.5 (1b), centers of cubes,  {\it B} at at 0.0 0.0 0.0 (1a), vertices of cubes,  and {\it O} at 0.5 0.0 0.0 (3d), midpoints of edges of cubes.}}
\label{fig:figure7}
\end{figure}

Figure~\ref{fig:figure7} shows two settings of the cubic {\it Pm$\bar{3}$m} perovskite structure, according to the origin of the lattice and the corresponding Wyckoff positions. This way of representing the perovskite structure by placing the origin on one of the two cations {\it A} or {\it B} corresponds to the representation by Miller and Love~\cite{millerlove}. 
\begin{table}[htbp!]
\renewcommand{\arraystretch}{1.35}
\centering
\caption{{\small Phonons modes label depending of the $A$ Wyckoff position in  {\it Pm$\bar{3}$m} reference high symmetry structure.}}
\begin{tabular}{lccc}
\hline                                             
\hline
\multicolumn{4}{c}{{\it A} ($\frac{1}{2}$, $\frac{1}{2}$, $\frac{1}{2}$) $\Longleftrightarrow$  {\it A} (0, 0, 0)} \\
{\it k}-points                                 & \multicolumn{3}{c}{Mode label}                  \\\hline
$\mathit{\Gamma}$ (0, 0, 0)                    & $\mathit{\Gamma}_4^- \Leftrightarrow \mathit{\Gamma}_4^-$ &$\mathit{\Gamma}_5^- \Leftrightarrow \mathit{\Gamma}_5^-$ & \\
$\Delta$ (0, $\frac{1}{4}$, 0)                 & $\Delta_5 \Leftrightarrow \Delta_5$ & &  \\
$X$ ($\frac{1}{2}$, 0, 0)                      & $X_5^+\Leftrightarrow X_5^-$ & $X_3^-\Leftrightarrow X_1^+$&   \\
$M$ ($\frac{1}{2}$, $\frac{1}{2}$, 0)          & $M_3^+\Leftrightarrow M_2^+$  & $M_3^-\Leftrightarrow M_2^-$  &$M_4^+\Leftrightarrow M_1^+$ \\
                                               & $M_5^+\Leftrightarrow M_5^+$ & $M_5^-\Leftrightarrow M_5^-$ & \\
$T$ ($\frac{1}{2}$, $\frac{1}{2}$, $\frac{1}{4}$)& $T_4\Leftrightarrow T_2$ & $T_5\Leftrightarrow T_5$ &  \\
$R$ ($\frac{1}{2}$, $\frac{1}{2}$, $\frac{1}{2}$)& $R_4^+\Leftrightarrow R_5^-$ & $R_5^+\Leftrightarrow R_4^-$&  \\
\hline
\hline
\end{tabular}
\label{tab:table7}
\end{table}
When analyzing phonon modes in ISODISTORD or AMPLIMODES, the modes labels  will depend on the Wyckoff specific setting used. On the main text, we have adopted the convention of setting the origin on the B-site cation, as shown in Figure  \ref{fig:figure7}-(b).  A dictionary between the labels of the different settings can be found in Table \ref{tab:table7}. 
\section{LDA energies of metastable phases}
\begin{figure*}[h!]
\centering
\begin{tabular}{cc}
 \includegraphics[width=5cm,keepaspectratio=true]{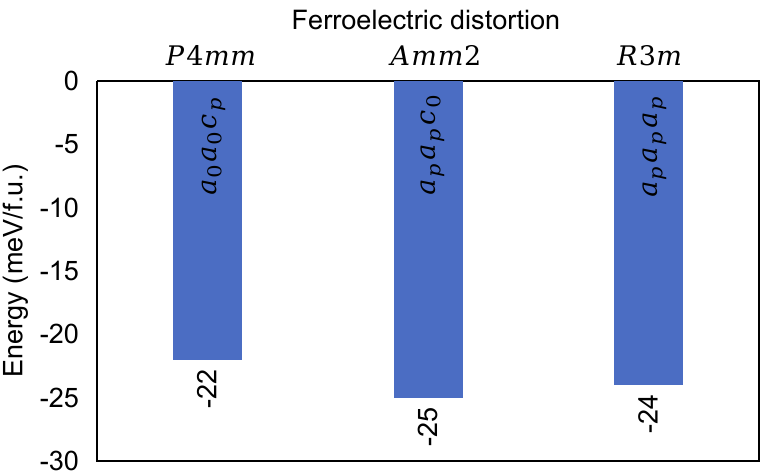} & \includegraphics[width=8cm,keepaspectratio=true]{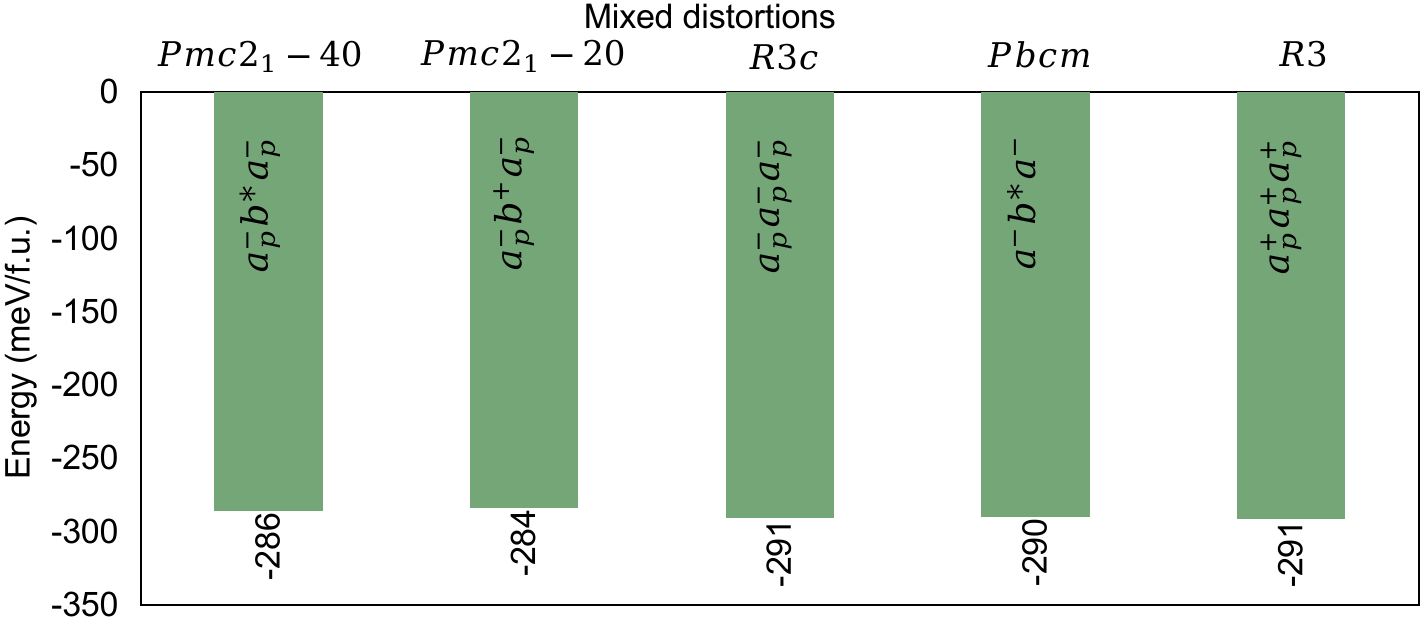}\\
 {\small (a)} & {\small (c)}\\
\multicolumn{2}{c}{\includegraphics[width=12.2cm,keepaspectratio=true]{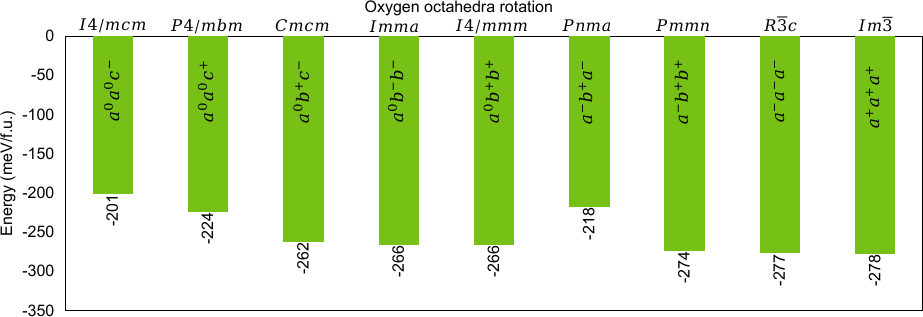} } \\
\multicolumn{2}{c}{{\small (b)}}   
\end{tabular}
\caption{{\small (Color online) Energy (meV/f.u.) of different fully relaxed phases of AgNbO$_3$, respect to the equilibrium cubic paraelectric {\it Pm$\bar{3}$m} state taken as reference, obtained using the LDA functional. The phases are labeled according to the same conventions as in Fig.~\ref{fig:figure3}. The energies of the $R3c$ ($-290.7$ meV/f.u.) and $R3$ ($-291.2$ meV/f.u.) phases are very close but $R3$ remains the ground state.}}
\label{fig:figure6}
\end{figure*}

Figure~\ref{fig:figure6} shows the energy relative to the high-symmetry cubic phase of the different phases of AgNbO$_3$ discussed in the main text, when working within the LDA approximation.
It can be observed that, regardless of the approximation used (LDA, PBE or PBEsol), the ground state is consistently found to be the {\it R$3$} phase as shown in Table~\ref{tab:compare}.
\begin{table}[htbp]
\centering
\caption{Total energies (meV/f.u.) respect to the cubic phase taken as reference, calculated using the GGA-PBE, GGA-PBEsol and LDA functionals with a plane-wave cutoff energy of 45 Ha. The supercell size (in unit $a_0$) and related $k$-point mesh are indicated for each case. The $R$3 phase is consistently found to be the most stable structure across all functionals. For PBE, the $Pmc2_1$-40 structure is not metastable and relaxes back to the $Pbcm$ structure; therefore, the reported PBE energy is identical to that of $Pbcm$ as highlighted by the asterisk \textsuperscript{*}.}
\begin{tabular}{lccccc}
\hline
Phase &Supercell & k-mesh & PBE& PBEsol& LDA \\
\hline
Pmc2$_1$-40 & $\sqrt{2}$$\times$$\sqrt{2}$$\times$4 &6$\times$6$\times$2  & -141.29\textsuperscript{*}&-198.48 &-285.72\\
{\it R$3$c} & 2$\times$2$\times$2 &4$\times$4$\times$4          & -141.84&-201.87 &-290.66\\
{\it Pbcm} & $\sqrt{2}$$\times$$\sqrt{2}$$ \times$4 &6$\times$6$\times$2        & -141.29\textsuperscript{*}&-202.48 &-290.45\\
{\it R$3$} & 2$\times$2$\times$2&4$\times$4$\times$4 & \textbf{-143.90}&\textbf{-204.57}&\textbf{-291.22} \\
\hline
\end{tabular}
\label{tab:compare}
\end{table}

\section{Structural data}
In this section, we present the structural properties of several metastable phases of AgNbO$_3$.
Tables \ref{tab:table5} and \ref{tab:table6} summarize the lattice parameters and mode amplitudes of these metastable phases, calculated using density functional theory (DFT) within the PBEsol approximation, and compare them with experimental data, when available.

\begin{table}[b]
\renewcommand{\arraystretch}{1.20}
\centering
\squeezetable
\caption{{\small Relaxed lattice parameters of some low energy phases of AgNbO$_3$ (PBEsol) and comparison to experimental data when available.}}
\begin{tabular}{lll}
\hline
\hline
                               & \multicolumn{2}{c}{Lattice parameters [\AA]{}} \\
               \cline{2-3}
 Phase&     Present   &Experimental               \\\hline

 {\it Pmc$2_1$}-40 & $a=$ 5.5263             & $a=$ 5.5520~\cite{yashima}                 \\
                & $b=$ 5.6015             & $b=$ 5.6091             \\
                & $c=$ 15.5113             & $c=$ 15.6477             \\
 {\it Pmc$2_1$}-20 & $a=$ 5.5409       &                \\
                & $b=$ 5.5798             &            \\
                & $c=$ 7.7738             &              \\
 {\it R$3$c}    & $a=$ 7.8342             & $a=$ 7.935~\cite{shigwada}                            \\
                               & $\alpha=$ 88.82$^\circ$ & $\alpha=$ 88.856$^\circ$               \\
 {\it Pbcm}     & $a=$ 5.5205  & $a=$ 5.5436~\cite{sciau}                \\
                & $b=$ 5.6009  & $b=$ 5.6071                    \\
                & $c=$ 15.5029 & $c=$ 15.565                  \\
{\it R$3$}      & $a=$ 7.8296  &                              \\
                              & $\alpha=$ 90.05$^\circ$ &             \\
\hline
\hline
\end{tabular}
\label{tab:table5}
\end{table}

\begin{table*}[b]
\renewcommand{\arraystretch}{1.20}
\centering
\squeezetable
\caption{{\small Contributions  of symmetry-adapted modes [\AA] to the atomic distortion of different metastable phases of AgNbO$_3$ respect to the cubic phase. Present theoretical results obtained using the PBEsol functional are compared to other values from the literature.}}
\begin{tabular}{cccccccc}
\hline
\hline
      &     \multicolumn{2}{c}{{\it Pbcm}}  & \multicolumn{4}{c}{{\it Pmc$2_1-40$}}                                     & {\it Pmn$2_1$}\\
 Mode &     Present     & Exp.~\cite{sciau} &  Present     & Theo~\cite{Moriwake}  &  Exp.~\cite{yashima} & Exp.~\cite{changszyf}  &     Present \\\hline
$R_4^+$    & 1.47 & 1.32     & 1.47 &1.25  &1.19  & 0.92   &0.81   \\
$T_4$      & 1.35 & 1.10    &1.27  &1.17      &0.93  & 0.57    &            \\
$\mathit{\Gamma}_4^-$ &      &      & 0.49 &0.61       &0.24  & 0.60   & 0.34\\
$\Delta_5$   &0.63 &  0.60 & 0.39 &0.31     & 0.48 & 0.43       &        \\
$T_5$      &      &      &0.17  & 0.20 &0.29  & 0.55         &       \\
$\mathit{\Gamma}_5^-$ &      &      &0.12 &0.16   &0.15  & 0.08   & 0.04  \\
$X_5^+$  &      &      &0.07  & 0.06 &0.06  &0.18    &0.16    \\
$M_3^+$  &      &      &0.05  &0.04  &0.30  & 0.58    &1.09    \\
$M_5^-$  &0.20 &0.15& 0.04 &0.03 & 0.14 & 0.32      &0.23          \\
$T_2$    &0.05 &0.02  &0.04  &0.03 &0.01   &  0.23  &              \\
$\Delta_1$ &      &      &0.03 &0.04  &0.07   & 0.19 &              \\
$R_5^+$   & 0.05 & 0.03     &0.03 & 0.04 &0.04    & 0.21    &0.01  \\
$X_3^-$   & 0.00 & 0.02 & 0.01 & 0.02 & 0.04 & 0.11   & \\
$M_2^+$   &      &      &0.00 & 0.00 & 0.08  & 0.50    & 0.01\\
$R_4^-$  &      &      &       &       &        &                   &0.05   \\
$X_4^-$   &      &      &       &       &        &                   & 0.02  \\
\hline
\hline
\end{tabular}
\label{tab:table6}
\end{table*}
\section{Dynamical stability of $Pbcm$ and $R3$ phases}
\label{sec:dynstability}
\begin{figure}[hb!]
\centering
\begin{tabular}{cc}
{\small (a)} & \includegraphics[width=7.4cm,keepaspectratio=true]{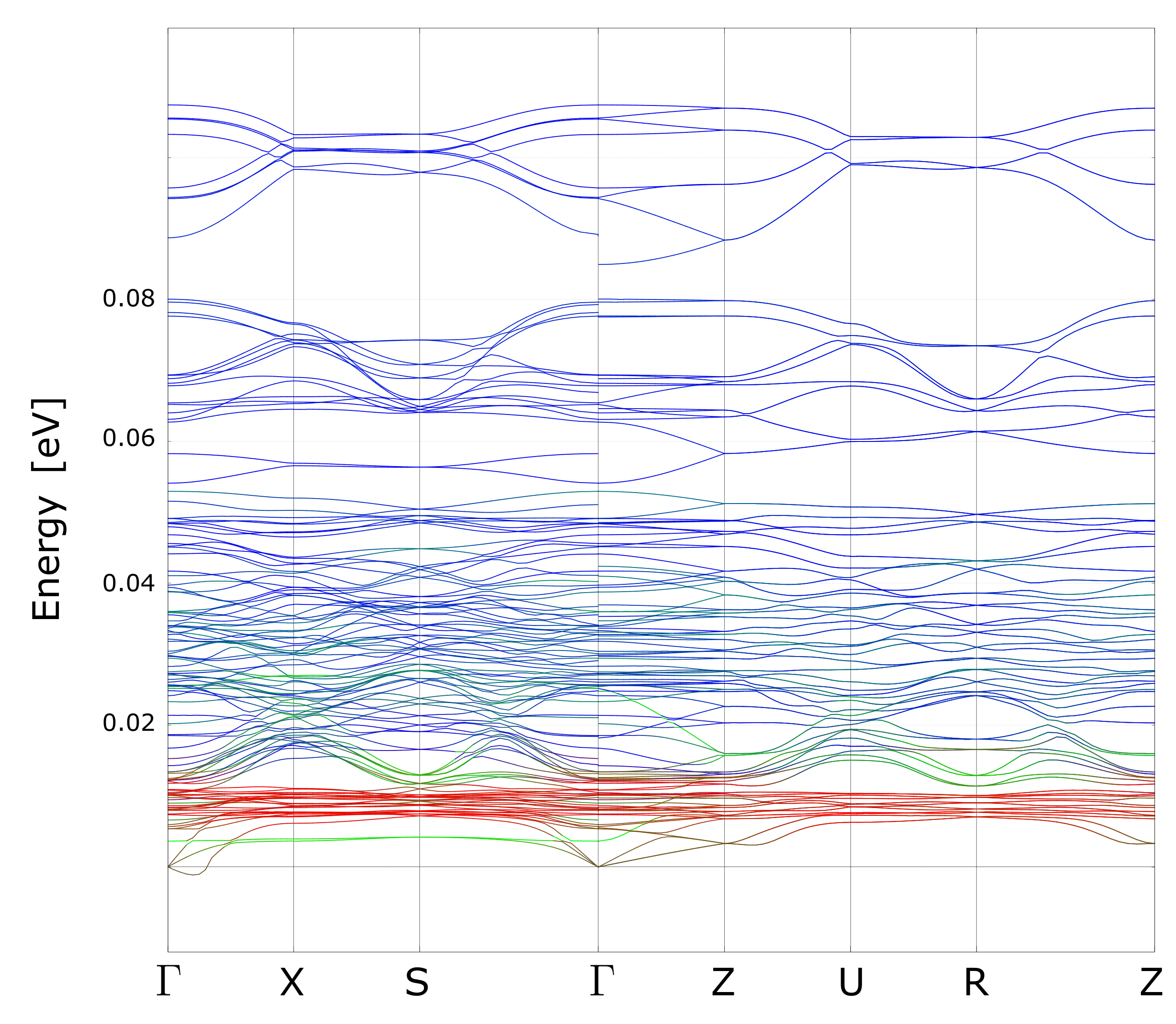} \\
{\small (b)} & \includegraphics[width=7.4cm,keepaspectratio=true]{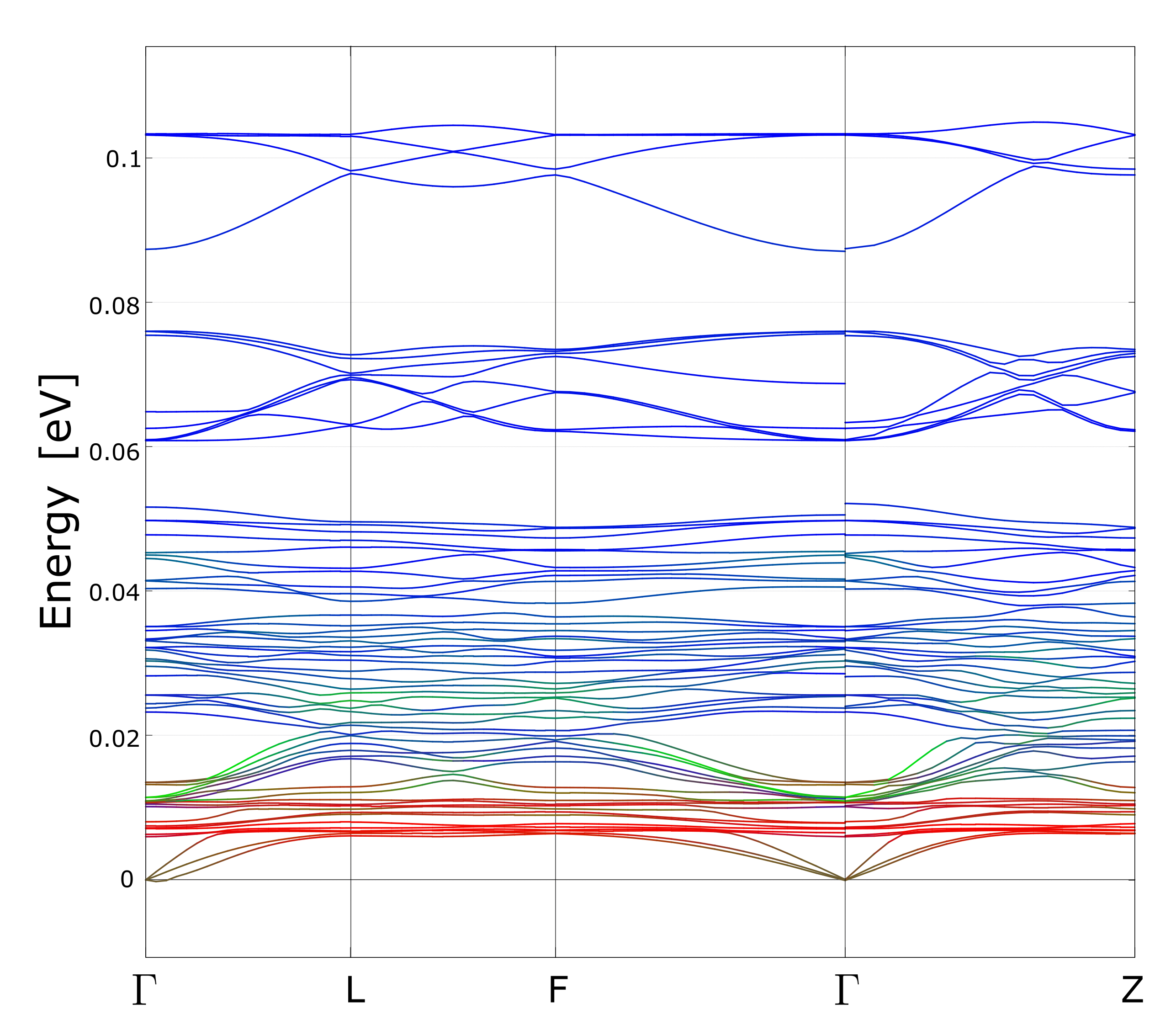} 
\end{tabular}
\caption{{\small (Color online) Calculated phonon dispersion curves of the (a)  $Pbcm$ phase of AgNbO$_3$ on the 40 atoms primitive cell using a $2\times2\times1$ $q$-point grid and (b) $R3$ phase of AgNbO$_3$ on the 20 atoms primitive cell using a $2\times2\times2$ $q$-point grid. A color is assigned along  each line, according to the contribution of each kind of atom to the associated eigendsplacement vector (red for the Ag/Na atom, green for the Nb atom, and blue for O atoms).}}
\label{fig:phononsPR}
\end{figure}
To assess the dynamical stability of the two lowest-energy structures identified in this work, we computed the phonon dispersions of the $R3$ and $Pbcm$ phases of AgNbO$_3$. The resulting phonon spectra are shown in Fig.~\ref{fig:phononsPR}. No imaginary phonon frequencies are found throughout the Brillouin zone for either structure, demonstrating that both phases correspond to local minima of the potential-energy surface and are dynamically stable at the harmonic level.

A small anomaly is nevertheless observed in the long-wavelength limit of the acoustic branches. Similar features have previously been reported in piezoelectric materials~\cite{Max-quadrupole} and were shown to originate from limitations of the standard phonon interpolation scheme when higher-order multipolar electrostatic contributions are neglected. To verify that this feature does not correspond to a genuine instability, we also computed the elastic tensors of both phases (see Table~\ref{tab:elastic}). In all cases, the elastic constants satisfy the mechanical stability criteria, confirming the anomaly in the long-wavelength limit of the acoustic branches as a technical artifact.
\begin{table}[H]
\caption{Elastic constants (in units of $10^2$ GPa) and minimal eigenvalue of the relaxed-ion elastic tensor for the $R3$ and $Pbcm$ phases of AgNbO$_3$. All eigenvalues are positive, confirming that both structures satisfy the mechanical stability criteria.}
\label{tab:elastic}
\centering
\begin{tabular}{lcc}
\hline\hline
& $R3$ & $Pbcm$ \\
\hline
$C_{11}$ & 2.1116 & 2.1114 \\
$C_{22}$ & 2.1116 & 1.9984 \\
$C_{33}$ & 2.1584 & 2.9690 \\
$C_{12}$ & 1.3334 & 1.1892 \\
$C_{13}$ & 1.1919 & 1.0695 \\
$C_{14}$ & -0.1593 &0.0000 \\
$C_{15}$ & -0.0034 &0.0000 \\
$C_{23}$ & 1.1919 & 1.2473 \\
$C_{44}$ & 0.2842 & 0.5116 \\
$C_{55}$ & 0.2842 & 0.4974 \\
$C_{66}$ & 0.3891 & 0.2309 \\
\hline
$\lambda_{\rm min}$ & 0.1688 & 0.2309 \\
\hline\hline
\end{tabular}
\end{table}

\begin{thebibliography}{58}
\bibitem [{pol()}]{polarization_note}%
  \BibitemOpen
  \href@noop {} {\bibinfo {title} {By default, polarization values reported in this work were computed using the {B}erry phase formalism; values in brackets are reported for comparison and correspond to estimates obtained within a linear approximation based on {B}orn effective charges and atomic displacements from the cubic reference phase}}\BibitemShut {NoStop}%
\end{thebibliography}
\end{document}